\documentclass[structabstract]{aa}  
\usepackage{graphicx}
\usepackage{txfonts}
\begin{document}
   \title{Ellipsoidal primary of the RS~CVn binary $\zeta$~And}

   \subtitle{Investigation using high-resolution spectroscopy and optical 
     interferometry\thanks{Based on observations collected at the European 
       Southern Observatory, Chile (Prg. 081.D-0216(AB)); and with the Vienna 
       automatic photometric telescopes Wolfgang and Amadeus, Arizona, USA}}

   \author{H. Korhonen\inst{1}
     \and
     M. Wittkowski\inst{1}
     \and
     Zs. K{\H o}v{\'a}ri\inst{2}
     \and
     Th. Granzer\inst{3}
     \and
     T. Hackman\inst{4}
     \and
     K.G. Strassmeier\inst{3}
   }

   \institute{European Southern Observatory, Karl-Schwarzschild-Str. 2,
     D-85748 Garching bei M{\"u}nchen, Germany\\
     \email{[hkorhone; mwittkow]@eso.org}
     \and
     Konkoly Observatory, H-1525 Budapest, PO Box 67, Hungary\\ 
     \email{kovari@konkoly.hu}
     \and
     Astrophysical Institute Potsdam, An der Sternwarte 16, D-14482 Potsdam, 
     Germany
     \and
     Observatory, PO Box 14, FI-00014 University of Helsinki, Finland}

   \date{Received; accepted}

 
  \abstract
  {}
  {We have obtained high-resolution spectroscopy, optical interferometry, and
    long-term broad band photometry of the ellipsoidal primary of the 
    RS~CVn-type binary system $\zeta$~And. These observations are used to 
    obtain fundamental stellar parameters and to study surface structures and 
    their temporal evolution.}
  {Surface temperature maps of the stellar surface were obtained from 
    high-resolution spectra with Doppler imaging techniques. These spectra were 
    also used to investigate the chromospheric activity using the H$\alpha$ 
    line and to correlate it with the photospheric activity. The possible 
    cyclicity in the spot activity was investigated from the long-term broad 
    band photometry. Optical interferometry was obtained during the same time 
    period as the high-resolution spectra. These observations were used to 
    derive the size and fundamental parameters of $\zeta$~And.}
  {Based on the optical interferometry the apparent limb darkened diameter of 
    $\zeta$~And is $2.55 \pm 0.09$ mas using a uniform disk fit. The expected 
    $\sim$4\% maximum difference between the long and short axes of the 
    ellipsoidal stellar surface cannot be confirmed from the current data which
    have 4\% errors. The Hipparcos distance and the limb-darkened diameter 
    obtained with a uniform disk fit give stellar radius of 
    $15.9\pm 0.8 {\rm R}_{\odot}$, and combined with bolometric luminosity, it 
    implies an effective temperature of $4665 \pm 140$~K. The temperature maps 
    obtained from Doppler imaging show a strong belt of equatorial spots and
    hints of a cool polar cap. The equatorial spots show a concentration around
    the phase 0.75, i.e., 0.25 in phase from the secondary, and another 
    concentration spans the phases 0.0--0.4. This spot configuration is 
    reminiscent of the one seen in the earlier published temperature maps of 
    $\zeta$~And. Investigation of the H$\alpha$ line reveals both prominences 
    and cool clouds in the chromosphere. These features do not seem to have a 
    clearly preferred location in the binary reference frame, nor are they 
    strongly associated with the cool photospheric spots. The investigation of 
    the long-term photometry spanning 12 years shows hints of a spot activity 
    cycle, which is also implied by the Doppler images, but the cycle length 
    cannot be reliably determined from the current data. }
   {}


   \maketitle

\section{Introduction}

Because of the enhanced dynamo action in stars with thick, turbulent 
outer-convection zones, rapidly rotating cool stars, both evolved and young,
exhibit significantly stronger magnetic activity than is seen in the Sun.
This activity means that the spots are also much larger than the spots observed
in the Sun. The largest starspot recovered with Doppler imaging is on the 
active RS~CVn-type binary HD 12545 which, in January 1998, had a spot that 
extended approximately 12$\times$20 solar radii (Strassmeier \cite{str99}). The 
lifetime of the large starspots/spot groups can also be much longer than that 
of the sunspots, even years instead of weeks for sunspots (e.g., Rice \& 
Strassmeier \cite{rice_str}; Hussain \cite{hussain}). The most typical dynamo 
signature is the presence of an activity cycle. Cyclic changes in the level of 
magnetic activity are well documented for the Sun, as well as for many 
solar-type stars (see, e.g., Ol{\'a}h et al.~\cite{cycles}). It is also 
interesting that, according to theoretical calculations, cyclic 
variations in the stellar magnetic activity can only be produced when 
differential rotation is present (R{\"u}diger et al.~\cite{ruediger}).

In this work \object{$\zeta$~Andromedae}, a long-period (17.8 day), single-lined
spectroscopic RS~CVn binary (Campbell~\cite{campbell}; Cannon~\cite{cannon}), 
is investigated in detail. In this system the primary is of spectral type 
K1~III, and the unseen companion possibly of type F (Strassmeier et 
al.\,\cite{strass93}). The primary fills approximately 80\% of its Roche lobe, 
so it has a non-spherical shape. The estimated ellipticity gives a $\sim$4\% 
difference between the long and short axes of the ellipsoid (K{\H o}v{\'a}ri 
et al.\,\cite{kovari07}, from here on Paper~1). The mean angular diameter of 
$\zeta$\,And has been derived to $2.72 \pm 0.036$\,mas using spectro-photometry
(Cohen et al. \cite{cohen})

An earlier  detailed Doppler imaging study (Paper~1) revealed that the spots on
the surface of $\zeta$~And have a temperature contrast of approximately 1000~K
and that they occur on a wide latitude range from the equator to an asymmetric
polar cap. The strength of the features changed with time, with the polar cap 
dominating the beginning of the two-month observing period in 1996/97, while 
the activity during the second half was dominated by medium-to-high latitude 
features. Also, the investigation revealed a weak solar-type differential 
rotation.

Here, results from Doppler imaging, optical interferometry, and long-term 
photometry of $\zeta$~And are presented. We discuss the reduction of the 
interferometric data and the obtained fundamental stellar parameters. The 
high-resolution spectra are used with Doppler-imaging techniques to obtain a 
surface temperature map. This surface map is compared to the earlier published 
temperature maps and also with the chromospheric activity based on observations
of the H$\alpha$ line. Finally, the long-term broad band photometry is used to 
study the temporal evolution of the spottedness, hence the possible spot cycles.

\section{Observations}

Simultaneous observations were carried out at the European Southern Observatory
with UVES (UV-Visual Echelle Spectrograph; Dekker et al. \cite{UVES}) mounted 
on the 8-m Kueyen telescope of the VLT, and the AMBER (Astronomical Multi BEam 
combineR; Petrov et al. \cite{AMBER}) instrument of the VLT Interferometer 
(VLTI). Additionally broad and intermediate band photometry in $V$, $Ic$ and 
$y$ bands were obtained with the automatic photoelectric telescopes Wolfgang 
and Amadeus in Arizona, USA (Strassmeier et al. \cite{kgs_APT}; Granzer et al. 
\cite{granz_APT}). For all the photometric observations HD~5516 was used as the
comparison star. 

All the observations were phased using the same ephemeris as in Paper~1,
\begin{displaymath}
{\rm HJD} = 2~449~997.223\pm0.017 + 17.769426\pm0.000040\times E,
\end{displaymath}
referring to the time of the conjunction.

\subsection{Optical interferometry}

\onltab{1}{
\begin{table*}
\caption{Log of the VLTI/AMBER observations.}
\label{tab:obsvlti}
\begin{tabular}{lllllllrrrr}
\hline\hline
Date & Time & Target & Purpose & FINITO & AM & Seeing & $\tau_0$\\
Sep. & UTC  &        &         &        &    & [$\arcsec$]&[msec]\\\hline

15   & 4:24-4:30 & $\theta$\,Psc & Calibrator     & on  & 1.17 & 0.66 & 3.1 \\
15   & 4:47-4:51 & $\theta$\,Psc & Calibrator     & off & 1.17 & 0.66 & 3.0 \\
15   & 5:04-5:07 & $\mu$\,Peg    & Check star     & on  & 1.63 & 0.65 & 3.1 \\
15   & 5:12-5:16 & $\mu$\,Peg    & Check star     & off & 1.65 & 0.62 & 3.2 \\
15   & 5:24-5:27 & 41\,Psc       & Calibrator     & on  & 1.19 & 0.67 & 2.9 \\
15   & 5:32-5:35 & 41\,Psc       & Calibrator     & off & 1.19 & 0.69 & 2.8 \\
15   & 5:45-5:49 & $\mu$\,Peg    & Check star     & on  & 1.79 & 0.62 & 3.1 \\
15   & 5:55-5:58 & $\mu$\,Peg    & Check star     & off & 1.85 & 0.54 & 3.6 \\
15   & 6:03-6:06 & $\mu$\,Peg    & Check star     & on  & 1.90 & 0.69 & 2.8 \\
15   & 6:14-6:17 & $\theta$\,Psc & Calibrator     & on  & 1.30 & 0.71 & 2.7 \\
15   & 6:22-6:25 & $\theta$\,Psc & Calibrator     & off & 1.33 & 0.62 & 3.1 \\
15   & 6:46-6:49 & $\zeta$\,And  & Science target & on  & 1.58 & 0.71 & 2.9 \\
15   & 6:53-6:57 & $\zeta$\,And  & Science target & off & 1.60 & 0.79 & 2.6 \\
15   & 7:02-7:06 & 41\,Psc       & Calibrator     & on  & 1.32 & 0.71 & 2.9 \\
15   & 7:13-7:22 & $\zeta$\,And  & Science target & on  & 1.66 & 0.65 & 3.2 \\
15   & 7:29-7:33 & HD\,7087      & Calibrator     & on  & 1.53 & 0.80 & 3.2 \\
15   & 7:38-7:41 & HD\,7087      & Calibrator     & off & 1.55 & 0.82 & 3.2 \\
15   & 7:53-7:56 & $\zeta$\,And  & Science target & on  & 1.84 & 0.97 & 2.2 \\
15   & 8:05-8:08 & HD\,15694     & Calibrator     & on  & 1.13 & 0.73 & 2.9 \\
\\
17   & 5:11-5:14 & $\theta$\,Psc & Calibrator     & on  & 1.19 & 0.91 & 2.4 \\
17   & 5:42-5:46 & $\mu$\,Peg    & Check star     & on  & 1.82 & 1.02 & 2.1 \\
17   & 6:08-6:15 & 41\,Psc       & Calibrator     & on  & 1.22 & 1.09 & 2.0 \\
17   & 6:36-6:40 & $\mu$\,Peg    & Check star     & on  & 2.30 & 0.97 & 2.5 \\
17   & 6:56-6:57 & $\zeta$\,And  & Science target & on  & 1.62 & 0.85 & 2.6 \\
17   & 7:09-7:13 & HD\,7087      & Calibrator     & on  & 1.50 & 0.86 & 2.5 \\
17   & 7:25-7:27 & $\zeta$\,And  & Science target & on  & 1.74 & 0.86 & 2.5 \\
17   & 7:41-7:44 & HD\,15694     & Calibrator     & on  & 1.12 & 0.91 & 2.4 \\
17   & 7:57-8:00 & $\zeta$\,And  & Science target & on  & 1.92 & 0.73 & 2.9 \\
\\
19   & 3:47-3:53 & $\theta$\,Psc & Calibrator     & on  & 1.18 & 0.74 & 1.9 \\
19   & 5:13-5:16 & $\mu$\,Peg    & Check star     & on  & 1.72 & 0.78 & 1.8 \\
19   & 5:28-5:31 & $\theta$\,Psc & Calibrator     & on  & 1.23 & 0.75 & 1.8 \\
19   & 5:46-5:49 & $\mu$\,Peg    & Check star     & on  & 1.90 & 1.06 & 1.3 \\
19   & 6:01-6:04 & 41\,Psc       & Calibrator     & on  & 1.22 & 0.97 & 1.4 \\
19   & 6:19-6:21 & $\mu$\,Peg    & Check star     & on  & 2.18 & 0.89 & 1.3 \\
19   & 6:54-6:57 & $\zeta$\,And  & Science target & on  & 1.64 & 0.81 & 1.7 \\
19   & 7:19-7:22 & HD\,7087      & Calibrator     & on  & 1.54 & 1.18 & 1.2 \\
19   & 7:52-7:55 & $\zeta$\,And  & Science target & on  & 1.95 & 1.55 & 0.9 \\
19   & 8:18-8:22 & HD\,15694     & Calibrator     & on  & 1.16 & 1.64 & 0.9 \\
\hline
\end{tabular}
\end{table*}}

The AMBER observations were obtained during the second part of the nights 
starting on September 14, 16, and 18, 2008, corresponding to orbital phases  
$\phi=0.05$ (secondary in front),  $\phi=0.15$ (intermediate case), and  
$\phi=0.27$ (secondary to the side), respectively. The details of the 
observations are listed in Table~\ref{tab:obsvlti}, only available online. 
During the night starting September 18 the coherence time was very short, so 
the data quality is lower than during the other half nights. For all the 
observations, AMBER was used in the low-resolution mode at $J$, $H$, and $K$ 
passbands, giving a resolving power ($\lambda/\Delta\lambda$) of $\sim 35$ 
and recording data between about 1.1-2.5\,$\mu$m. Only the $H$ and $K$ band 
data ($\sim$\,1.5-2.5\,$\mu$m) were used for the data analysis. The $J$ band 
data were of poor quality owing to vanishing detected flux. The fringe tracker 
FINITO (Le Bouquin et al. \cite{FINITO}) was used for most observations. During
the night starting September 14, data were also taken without the use of FINITO
in order to confirm the calibration of the visibility. The Auxiliary Telescopes
(ATs) were placed at the stations A0, K0, and G1, giving ground-baseline 
lengths of 128m (A0-K0) and 90m (A0-G1 and K0-G1). The A0-G1 and K0-G1 
baselines have the same ground length, but differ in position angle by 
90\,$\deg$. 

In addition to $\zeta$~And, a circular check star was observed every night. For
this \object{$\mu$~Peg} was chosen because it is at a similar position on the 
sky as $\zeta$~And and it is expected to have a similar angular diameter 
($\Theta_\mathrm{LD}$=2.50\,$\pm$\,0.04\,mas; Nordgren et al. \cite{nordgren}; 
Mozurkewich et al. \cite{mozurkewich}). Observations of $\zeta$~And and 
$\mu$~Peg were interleaved with observations of the interferometric calibration
stars $\theta$\,Psc (K1\,III, $K$=1.86, 
$\Theta_\mathrm{LD}$=2.00 $\pm$ 0.02\,mas), 41\,Psc (K3\,III, $K$=2.43, 
$\Theta_\mathrm{LD}$=1.81 $\pm$ 0.02\,mas), HD\,7087 (G9\,III, $K$=2.48, 
$\Theta_\mathrm{LD}$=1.59 $\pm$ 0.02\,mas), and HD\,15694 (K3\,III, $K$=2.48, 
$\Theta_\mathrm{LD}$=1.77 $\pm$ 0.02\,mas). The angular diameters for 
$\theta$\,Psc and 41\,Psc are from Bord{\'e} et al. (\cite{borde}) and those 
for HD\,7087 and HD\,15694 are from M{\'e}rand et al. (\cite{merand}).

\subsection{Spectroscopy}

The UVES observations of $\zeta$~And were carried out during 10 nights between 
September 13, 2008 and October 1, 2009. The red arm in the standard wavelength 
setting of 600~nm was used with the imageslicer~\#3. This instrument setup 
gives a spectral resolution ($\lambda/\Delta\lambda$) of 110\,000 and a 
wavelength coverage of 5000-7000 {\AA}. Each observation consists of three 
exposures of 8 seconds that were later combined to one very high 
signal-to-noise ratio (S/N) spectrum. The S/N of combined observations was 
between 586 and 914 around 6400\,{\AA}. The data were reduced using the UVES 
pipeline. A summary of the spectroscopic observations is given in the on-line 
Table~\ref{table_UVES}.

\onltab{2}{
\begin{table*}
\caption{The high-resolution spectroscopy with UVES at VLT.}
\label{table_UVES}
\centering        
\begin{tabular}{c c c c}
\hline\hline            
Date & HJD & phase & S/N \\ 
     & 2450000+ &  & \\
\hline                        
13.09.2008 & 4722.81672 & 0.940 & 866 \\
15.09.2008 & 4724.70463 & 0.046 & 650 \\
17.09.2008 & 4726.71012 & 0.159 & 586 \\
18.09.2008 & 4727.68700 & 0.214 & 628 \\
19.09.2008 & 4728.73866 & 0.273 & 762 \\
21.09.2008 & 4730.72025 & 0.384 & 693 \\
23.09.2008 & 4732.72873 & 0.497 & 914 \\
25.09.2008 & 4734.65166 & 0.606 & 785 \\
27.09.2008 & 4736.69360 & 0.721 & 612 \\
01.10.2008 & 4740.75333 & 0.949 & 802 \\
\hline 
\end{tabular}
\end{table*}
}

\section{Reduction and analysis of the interferometric data}

\subsection{Data reduction}

Raw visibility and closure phase values were computed using the latest 
version of the {\tt amdlib} data reduction package (version 2.2) and the 
{\tt yorick} interface, both provided by the Jean-Marie Mariotti Center 
(JMMC). The data reduction principles are described in Tatulli et al. 
(\cite{tatulli}). 

Absolute wavelength calibration was performed by correlating the raw spectra 
with a model of the atmospheric transmission, resulting in a correction of 
$\Delta\lambda/\lambda=-0.043$ in the K-band with respect to the original 
wavelength table (cf. Wittkowski et al. \cite{witt08}). For each observation 
mentioned in Table~\ref{tab:obsvlti}, only some of the individual frames were 
selected for further analysis. Only those frames were used that had a flux 
ratio under 3 between the telescopes of the concerned baseline and that had an 
estimated absolute piston of less than 4\,$\mu$m. Finally, out of these only 
the 30\% of the frames with the highest fringe signal-to-noise ratio were kept.
The selected frames were averaged.

The resulting differential phase and visibility values were significantly 
affected by chromatic piston effects caused by the dispersion of the air 
(cf. Millour et al. \cite{millour}; Le Bouquin et al. \cite{lebouquin}). 
This effect was relatively strong for our data because of the combination of 
long baselines and large airmasses. We used the measured differential phase to 
estimate the amount of chromatic piston $\delta$ using $\delta=1/2\pi\,\, 
d\phi/d\sigma$, where $\phi$ is the differential phase and $\sigma$ the 
wavenumber. The loss of the squared visibility amplitude $\rho$ was estimated 
using formula (1) of Millour et al. (\cite{millour}). The averaged visibility 
data were compensated using the estimated $\rho$. Millour et al. note that 
$\delta$ is the absolute piston value relative to the white light fringe. The 
absolute piston also includes the frame-by-frame piston that is estimated by 
the regular AMBER data reduction. This quantity is determined with respect to 
the pixel-to-visibility matrix (P2VM) reference, which can have an offset to 
the white light fringe. We selected frames with an estimated piston of less 
than 4\,$\mu$m, and verified that the piston of our P2VM measurements is less 
than 2\,$\mu$m in the $H$ band and less than 4\,$\mu$m in the $K$ band. In 
total, we assumed an error of the piston estimate of 5\,$\mu$m and propagated 
it to the final visibility amplitude. We also used an alternative compensation 
of the loss of the squared visibility amplitude $\rho$ that was based on a 
parametrization of the calibrator star data as a function of optical path 
difference, i.e., an estimate that does not depend on the measured differential
phase of the science target. We obtained results well within the adopted error.

As a final data reduction step, the squared visibility amplitudes were 
calibrated for the interferometric transfer function, which was estimated 
using an average of the computed transfer functions based on the closest 
calibration star measurement before that of each science target and the 
closest thereafter. The final error of the calibrated data includes the
statistical error of the frames, the error in the correction for chromatic 
piston, and the standard deviation of the two transfer function measurements.

\begin{figure}
  \centering
  \includegraphics[width=8cm]{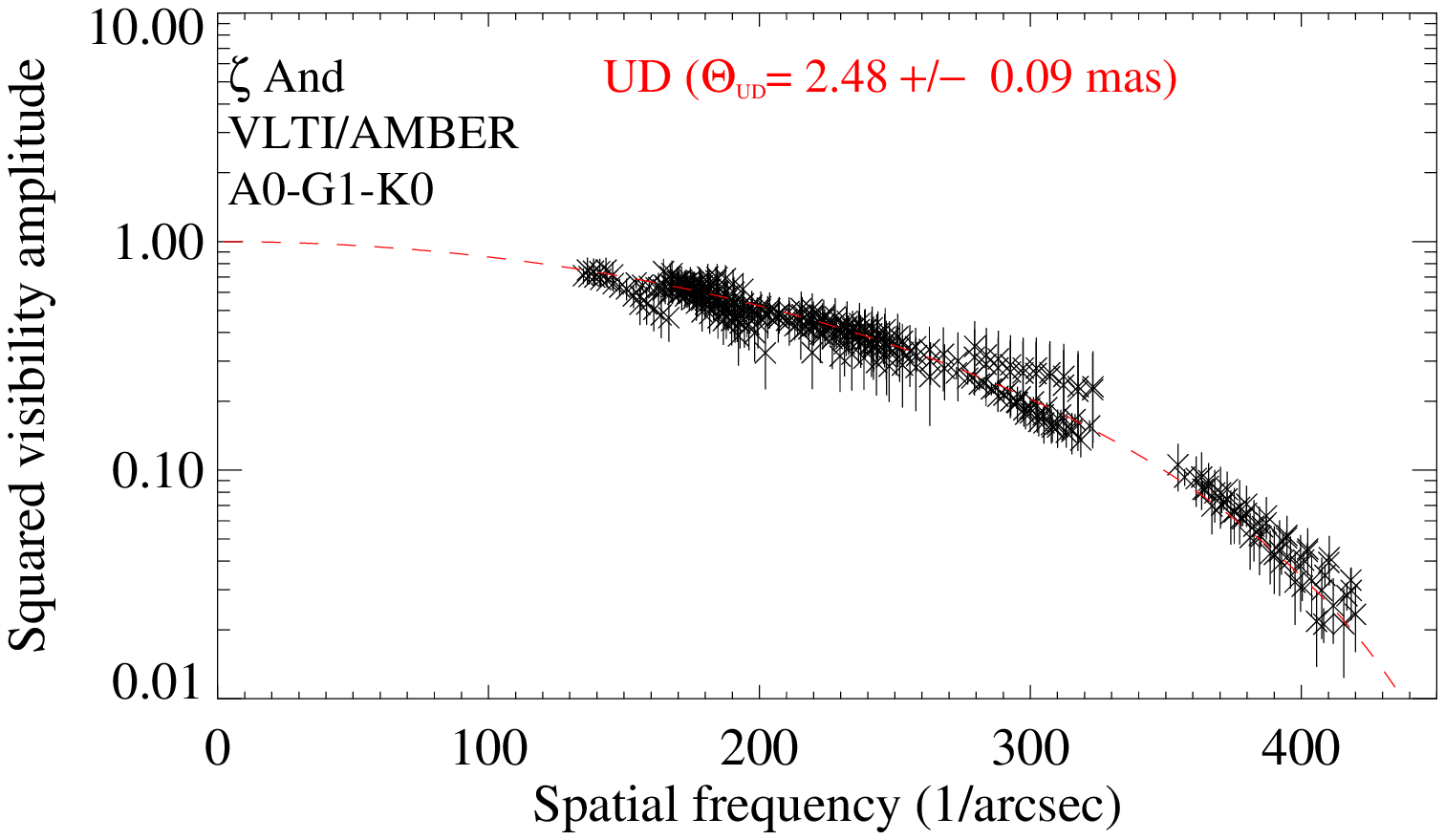}
  \includegraphics[width=8cm]{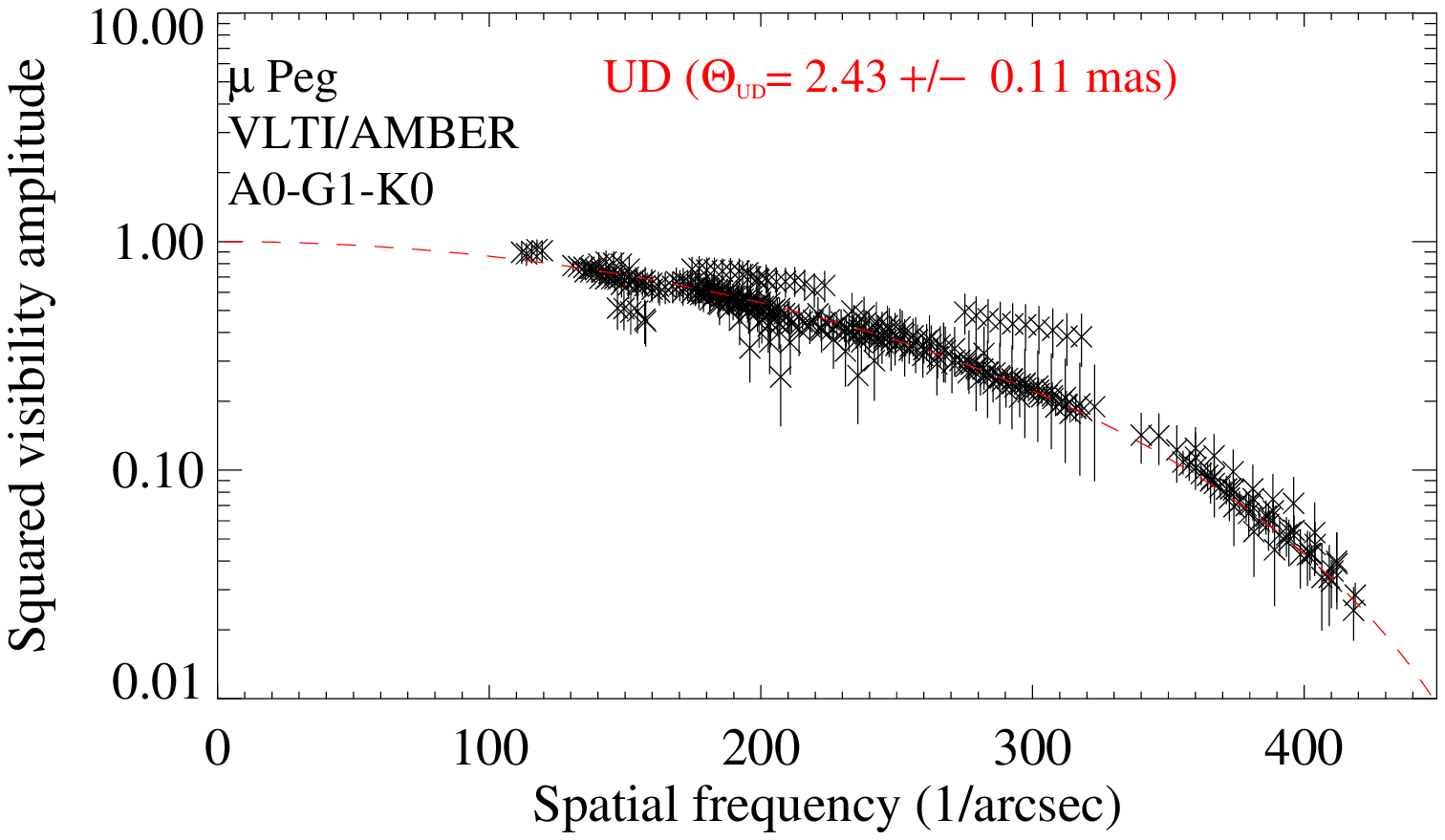}
  \caption{VLTI/AMBER visibility data of $\zeta$~And (top) and of the check 
    star $\mu$~Peg (bottom), compared to uniform disk models.}
  \label{fig:vlti}
\end{figure}

\subsection{Analysis of interferometric data}

\begin{table}
\caption{Uniform disk fit results for the VLTI/AMBER data.}
\label{tab:resvlti}
\centering
\begin{tabular}{ll|rr}
\hline\hline
Day  & FINITO & $\zeta$\,And & $\mu$\,Peg \\\hline
14   & ON     & 2.48 $\pm$ 0.06 mas & 2.43 $\pm$ 0.07 mas \\
14   & OFF    & 2.40 $\pm$ 0.21 mas & 2.43 $\pm$ 0.06 mas \\
16   & ON     & 2.53 $\pm$ 0.11 mas & 2.42 $\pm$ 0.19 mas \\
18   & ON     & 2.58 $\pm$ 0.12 mas & 2.62 $\pm$ 0.16 mas \\\hline
$\theta_\mathrm{UD}$ & all data  & 2.48 $\pm$ 0.09 mas & 2.43 $\pm$ 0.11 mas 
\\[1ex]
$\theta_\mathrm{LD}$ & all data  & 2.55 $\pm$ 0.09 mas & 2.49 $\pm$ 0.11 mas 
\\ \hline
\end{tabular}
\end{table}

Figure~\ref{fig:vlti} shows the resulting visibility data of $\zeta$~And and of 
the check star $\mu$~Peg obtained from all three observing nights and compared 
to models of a uniform disk (UD). Because of the relative large errors in the 
observations, no differences were seen in the measurements from different 
baselines. Thus all the baselines were used together in the analysis. 
Table~\ref{tab:resvlti} lists the resulting uniform disk diameters of 
$\zeta$~And and $\mu$~Peg for each of the nights separately, as well as for all
three observing nights. When using the data from all the nights together, the 
diameter is estimated from the data obtained during the nights starting on 
September 14 and 16, but the error is from all the data, i.e., including the 
data from the night starting on September 18. This is done because the data 
quality is significantly lower on the night starting on September 18 than 
during the two other observing nights. 

During the night of September 14, we obtained data with and without the use of 
the fringe tracker FINITO. The results for these two data sets agree well 
within the errors for both targets, and we do not see any systematic 
calibration effects that are caused by the use of FINITO. Deviations between 
observed visibility values and the UD model are mostly caused by residuals of 
the compensation of the chromatic piston effect, which was most noticeable on 
the baseline A0-G1, and by systematic calibration uncertainties due to varying 
atmospheric conditions. Within the obtained errors of the UD diameter of about 
4\%, we do not see indications of any elliptical intensity distribution of 
$\zeta$~And. However, the ellipticity of $\sim$4\% expected for the night of 
September 18 is consistent with our data. 

Correction factors between UD diameter and limb-darkened (LD) disk diameters 
were computed using ATLAS\,9 model atmospheres (Kurucz \cite{kurucz}). For the 
spectral types of our target stars $\zeta$~And and $\mu$~Peg and the wavelength
range used for our observations, we obtain values for 
$\theta_\mathrm{UD}/\theta_\mathrm{LD}$ of 0.974 and 0.976, respectively. The 
resulting LD diameters are $\theta_\mathrm{LD} =2.55 \pm 0.09$\,mas and 
$\theta_\mathrm{LD} =2.49 \pm 0.11$\,mas, respectively. Cohen et al. 
(\cite{cohen}) give a diameter of $2.72\pm 0.036$\,mas for $\zeta$\,And, based 
on spectro-photometry. This diameter is significantly larger, but the error 
smaller, than what was obtained in this work. Still, the spectro-photometric 
observations could be affected by the significant magnetic activity exhibited 
by $\zeta$~And. The LD diameter of $\mu$~Peg obtained here is consistent 
with the earlier interferometric measurements of $\theta_\mathrm{LD} =2.53 \pm 
0.09$\,mas obtained with the NPOI and $\theta_\mathrm{LD} =2.49 \pm 0.04$\,mas 
obtained with the Mark\,III interferometers (Nordgren et al. \cite{nordgren} 
and Mozurkewich et al. \cite{mozurkewich}), increasing the confidence in the 
results presented here. 

\section{Fundamental parameters}

\subsection{Radius}

The limb darkened diameter of $\zeta$~And, obtained from the interferometric 
observations, is $2.55\pm 0.09$ mas. Together with the Hipparcos parallax of 
$17.24\pm 0.26$ mas (van Leeuwen \cite{hip}) this can be used to determine the 
stellar radius with the following formula: 
$R = \Theta_{\rm LD}\frac{C}{2\pi_{\rm p}}$, where $\Theta_{\rm LD}$ is the 
limb-darkened angular diameter in radians, $\pi_{\rm p}$ the parallax in 
arcseconds, and $C$ the conversion from parsecs to meters. For $\zeta$~And 
this gives stellar radius of $R=15.9\pm 0.8 {\rm R}_{\odot}$, which is 
consistent with the 16.0~R$_{\odot}$ estimated in Paper\,1.

\begin{figure*}
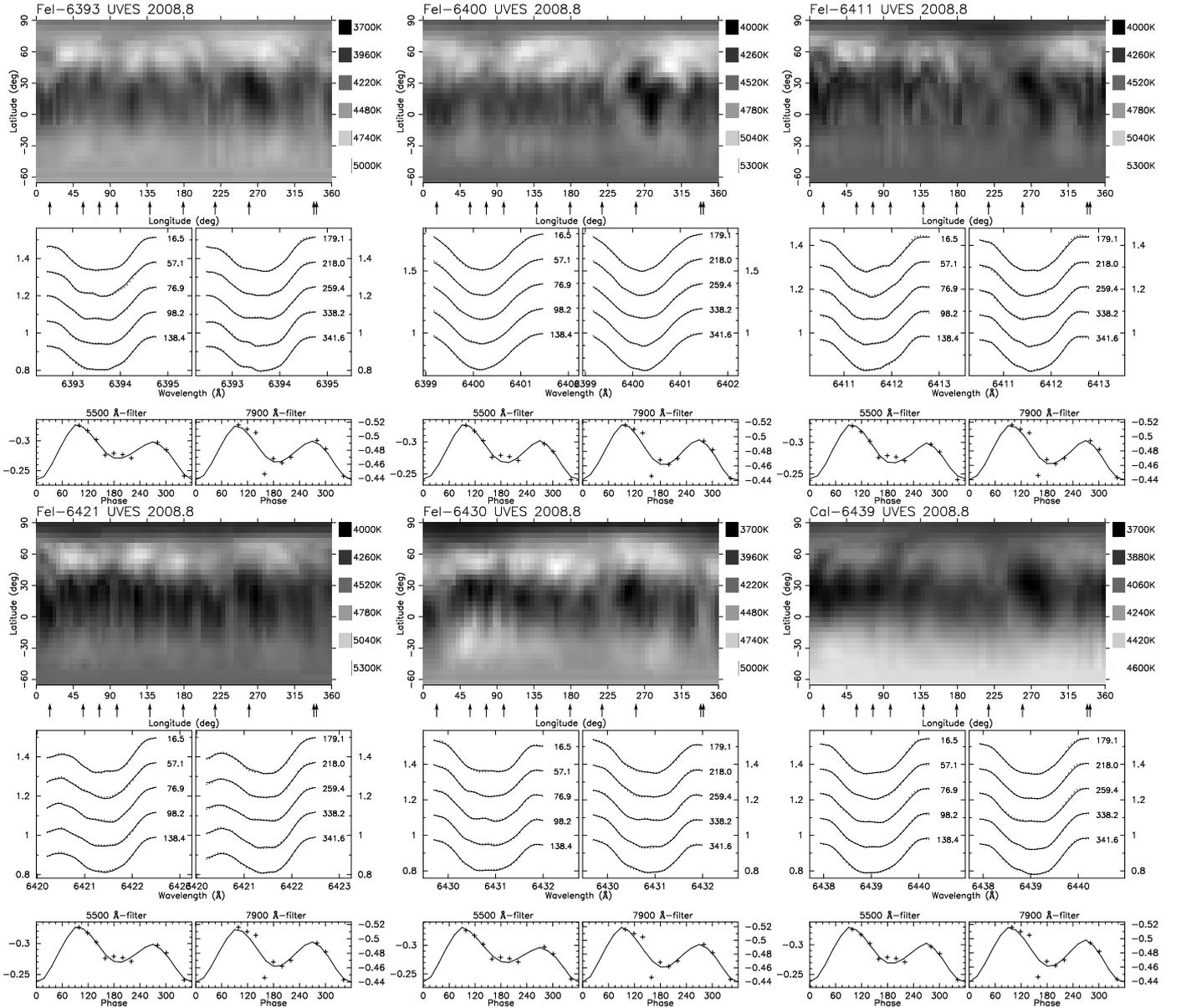

  \centering
  \includegraphics[width=6cm]{13736fg2a.ps}
  \includegraphics[width=6cm]{13736fg2b.ps}
  \includegraphics[width=6cm]{13736fg2c.ps}
  \includegraphics[width=6cm]{13736fg2d.ps}
  \includegraphics[width=6cm]{13736fg2e.ps}
  \includegraphics[width=6cm]{13736fg2f.ps}
  \caption{Doppler imaging results of $\zeta$~And obtained with the 
    {\sc TempMap}$_\epsilon$ code. In each plot the temperature map for an 
    individual spectral line is shown with the observed spectra and 
    photometry, including the corresponding fits to the data. In the middle 
    panels, under the continuous line fits, the tiny vertical dashes
    are the 1-$\sigma$ error bars of the spectroscopic observations.}
  \label{doppler_maps}
\end{figure*}

\subsection{Effective temperature}

The effective temperature of a star can be calculated from the interferometric 
diameter determination when combined with a bolometric flux measurement using 
the formula
\begin{equation}
T_{\rm eff}=\left(\frac{4f_{\rm bol}}{\sigma\Theta_{\rm LD}^{2}}\right)^{1/4},
\label{eq:Teff}
\end{equation}
where $f_{\rm bol}$ is the bolometric flux and $\sigma$ is the Stefan-Boltzmann
constant.

The bolometric flux of $\zeta$~And was estimated using measurements on all
the available photometric passbands and the {\tt getCal} tool of the NASA 
Exoplanet Science Institute's interferometric observation planning tool suite. 
The bolometric flux of $f_{\rm bol} = 9.863 \pm 0.54 \times 10^{-10}$~W/m$^2$ 
was obtained. Inserting this value to Eq.~\ref{eq:Teff}, together with the 
limb-darkened angular diameter, gives $T_{\rm eff}$ of $4665\pm 140$~K. This 
value is very close to, and the same within errors, as the 
$T_{\rm eff}\approx 4600$\,K used for Doppler imaging in Paper\,1 and the 
current work.

\section{Doppler imaging}

For Doppler imaging we used the code {\sc TempMap}, which was originally written
by Rice et al. (\cite{rice89}). The code performs a full LTE spectrum synthesis
by solving the equation of transfer through a set of ATLAS-9 (Kurucz 
\cite{kurucz}) model atmospheres at all aspect angles and for a given set of 
chemical abundances. Simultaneous inversions of the spectral lines, as well as 
of the two photometric bandpasses, are then carried out using a maximum-entropy 
regularization. For the non-spherical $\zeta$\,And, a new version of the code
was applied: {\sc TempMap}$_\epsilon$ (see Paper~1 and the references therein) 
takes the distorted geometry of the evolved component in a close binary into 
account through the distortion parameter $\epsilon$. The elliptical distortion 
is approximated by a rotation ellipsoid, elongated towards the secondary star:
$\epsilon = \sqrt{1 - \left(\frac{b}{a}\right)^2}$ 
where $a$ and $b$ are the long and the short axes of the ellipsoid, 
respectively. The appropriate value of $\epsilon$ for $\zeta$\,And, 0.27, as 
well as the overall system and stellar parameters, were adopted from Paper~1 
(Table~2 therein).

The 30 available UVES spectra (three exposures per night) covered 18 days, 
i.e., one full rotation cycle, thus allowing one Doppler reconstruction. The 
three nightly observations were averaged, since they were taken within 
approximately 120\,sec. Thus, for further investigation we used the ten 
averaged spectra with an enhanced S/N value of $\sim$600 or more (see 
Table~\ref{table_UVES} for more details on observations).

Doppler imaging was performed for the well-known mapping lines within the 
6392--6440~\AA\ spectral range. Doppler maps for Fe\,{\sc i}~6393, 6400, 6411, 
6421, 6430, and Ca\,{\sc i}~6439 are shown in Fig~\ref{doppler_maps}. The 
individual maps revealed similar spot distributions, i.e., mainly cool spots 
at low latitudes with temperature contrasts of 600--900\,K with respect to the 
unspotted surface of 4600\,K. Cool polar features are also recovered, however, 
with significantly weaker contrast ranging from $\sim$100\,K (Fe\,{\sc i}~6400)
to a maximum of $\sim$700\,K (cf. the Fe\,{\sc i}~6430 map). Numerous bright 
features also appear in the iron maps; however, as they occur near dominant 
cool spots they can be artifacts, so-called ``rebound'' features (see, e.g., 
Rice \cite{rice}).

\begin{figure}
  \centering
  \includegraphics[width=8cm]{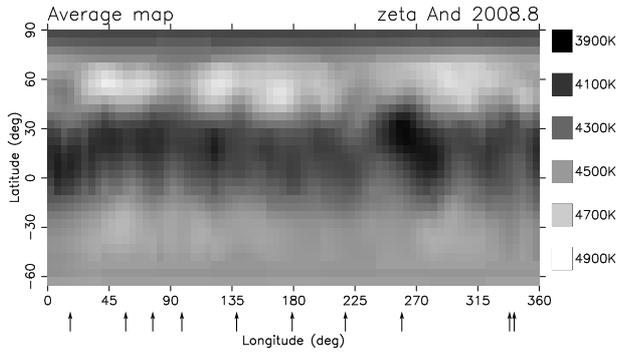}
  \caption{Average map produced from all the maps shown in 
    Fig.~\ref{doppler_maps}. }
  \label{averagemap}
\end{figure}

Despite the small difference between the temperature contrasts of the 
respective maps and the spurious bright features, the resulting six Doppler 
maps are in very good agreement. This similarity is more conspicuous in 
Fig~\ref{averagemap}, where the average of the six individual maps is plotted. 
Averaging did not blur the overall structure. The most prominent feature is the
belt of cool spots at the equatorial region, with the strongest concentration 
of spots located at the phase $\phi=$0.75 and at another cool region ranging 
between phases $\phi=0.0-0.4$. Also a weak polar feature can be detected. This 
result is reminiscent of the result in Paper~1, where low-latitude dominant 
features also tended to concentrate at quadrature positions of opposite 
hemispheres for both observing seasons.

\section{Discussion}

\subsection{Comparison between spherical and ellipsoidal surface geometry in 
the inversion}

Another Doppler imaging code, INVERS7PD, which was written by N.\ Piskunov 
(see, e.g., Piskunov et al.~\cite{pisk90}) and modified by T.\ Hackman (Hackman
et al.~\cite{hack}), was also used to obtain a temperature map of $\zeta$~And. 
In this inversion spherical geometry and only the Fe\,{\sc i}~6400~{\AA} line 
were used. The observations are compared to a grid of local line profiles 
calculated with the SPECTRUM spectral synthesis code (Gray \& Corbally 
\cite{SPECTRUM}) and Kurucz model atmospheres (Kurucz \cite{kurucz}). In the 
calculations, 10 limb angles and nine temperatures between 3500~K and 5500~K 
were used. Photometry was not used as a constraint in this inversion as the
ellipticity effect seen in the light curves cannot be properly taken into 
account when using spherical geometry.

\begin{figure}
  \centering
  \includegraphics[width=8cm]{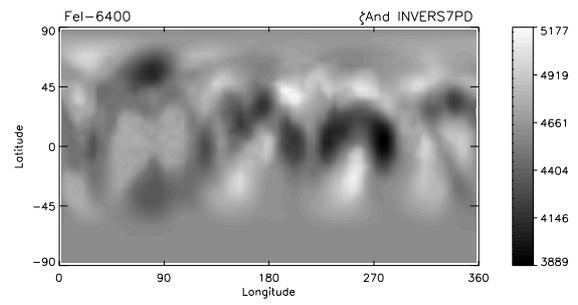}
  \caption{Temperature map of $\zeta$~And obtained with INVERS7PD inversion
code using spherical geometry and Fe\,{\sc i} 6400~{\AA} line.}
  \label{INVERS7}
\end{figure}

The result of this inversion is shown in Fig.~\ref{INVERS7}. The main spot 
structures and the temperature range are similar to the ones in the map 
obtained with {\sc TempMap}$_\epsilon$ from the Fe\,{\sc i} 6400~{\AA} line. 
The cool spots concentrate on the equatorial region and especially on the 
quadrature points. The main spot is seen at the phase $\phi$=0.75 in the 
equatorial region, and there is also a prominent spot at the phase $\phi$=0.25 
at higher latitudes. This is missing from the map obtained using 
{\sc TempMap}$_\epsilon$, so it is most likely an artifact caused by 
using spherical geometry on an ellipsoidal star (cf. Fig.3 in Paper 1). 
Furthermore, at the phase $\phi$=0.25, the equatorial region has a temperature 
close to that of the unspotted surface, unlike in the map obtained using 
ellipsoidal geometry. Also, the whole temperature scale is shifted by 100~K 
towards the cooler temperatures. On the whole, the temperature map obtained 
with the spherical geometry is very similar to the one obtained using 
ellipsoidal geometry and {\sc TempMap}$_\epsilon$. As expected, the main 
differences occur at the quadrature points and especially at the phase 
$\phi$=0.25. Also, one has to keep in mind that the tests with 
{\sc TempMap}$_\epsilon$ show that neglecting the ellipticity in the Doppler 
imaging reconstruction yields $\sim$50--240\% higher $\chi^{2}$ values in 
comparison to using the correct surface geometry.

\begin{figure}
  \centering
  \includegraphics[width=7cm]{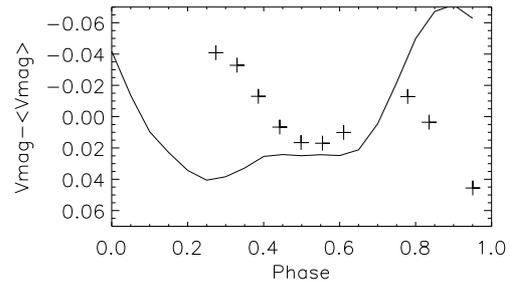}
  \caption{Observed $V$ magnitude of $\zeta$~And (crosses) compared to the
    one calculated from the temperature map obtained using INVERS7PD (line).}
  \label{INVERS7_phot}
\end{figure}

The main differences between the results from the spherical and ellipsoidal 
codes can be seen in the photometry. Figure~\ref{INVERS7_phot} shows the 
normalised $V$ observations of $\zeta$~And compared to the ones calculated from
the temperature map obtained with the code using spherical geometry. Only 
around the phases 0.4--0.6 the two light curves show similar behaviour, 
and the photometry calculated from the INVERS7PD temperature map shows 
completely different behaviour than the observed one especially around the 
quadrature points. 

\subsection{Chromospheric activity}

\begin{figure}
  \centering
  \includegraphics[width=8cm]{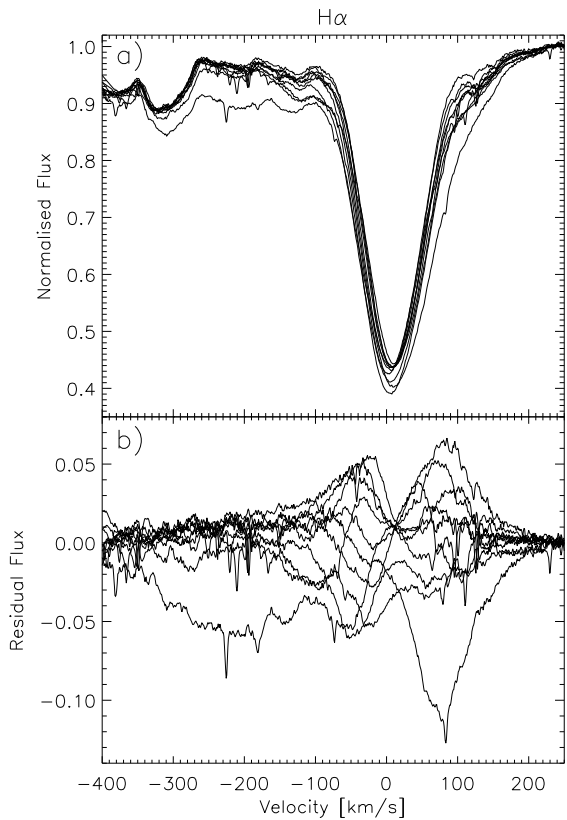}
  \caption{Variations in the H$\alpha$ profiles of $\zeta$~And. a) The 
    continuum normalised H$\alpha$ profiles. b) Residual profiles after the 
    mean profile has been subtracted. The x-axis gives the velocity in 
    comparison to the mass centre of the giant component.}
  \label{Ha_prof}
\end{figure}

Chromospheric activity of $\zeta$~And was investigated using the H$\alpha$ line
profiles, which appeared in absorption during the observations, similarly to 
the other Balmer lines. Variations in the H$\alpha$ line through the rotation 
cycle are shown in Fig.~\ref{Ha_prof}a. Both the red and the blue wings show 
strong variation at one, but different, phase. Also, most line profiles clearly
show variable behaviour between velocities -100~km/s and +100~km/s. These 
variations are clearly seen already in the spectra, which have not been 
corrected to the continuum level. All the spectra show identical continuum 
shapes, except approximately $\pm$5{\AA} from the H$\alpha$ line, corresponding
to the variation also seen in the normalised spectra used in the following 
analysis.

\begin{figure}
  \centering
  \includegraphics[width=8cm]{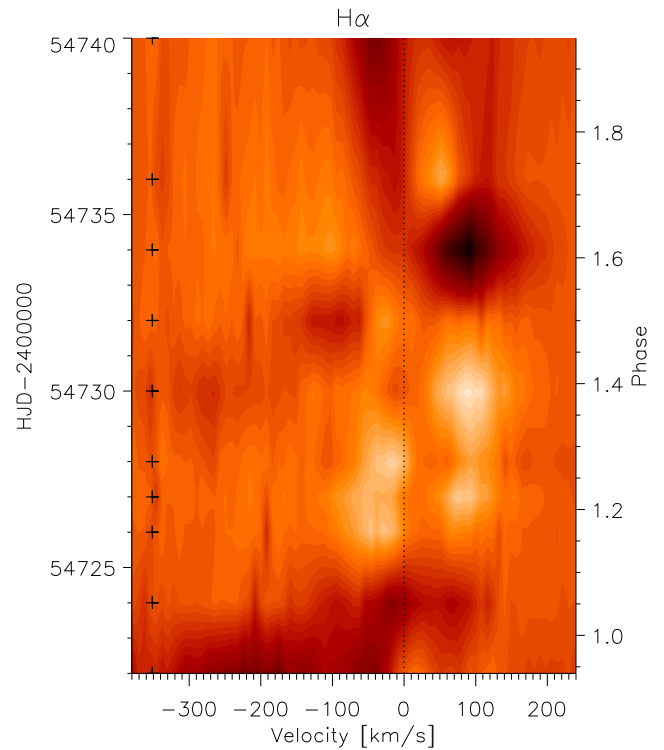}
  \caption{Dynamic spectrum of the H$\alpha$ line based on the residual 
    profiles after the subtraction of the average line profile from the 
    individual profiles. The x-axis gives the velocity and the y-axis the 
    heliocentric Julian date on the left and phase on the right side. The 
    crosses on the lefthand side of the plot give the heliocentric Julian 
    dates of the observations and the dotted line the zero velocity. The 
    brighter the colour, the more emission is observed.}
  \label{Ha_hjd}
\end{figure}

To investigate the line-profile variations in more detail, the average profile 
was subtracted from all the profiles, thus creating the residual profiles shown
in Fig.~\ref{Ha_prof}b. The temporal variations are clearly seen in these 
profiles. The most prominent features are the two strong absorption features 
seen at the velocities -350~-- 0~km/s and 0~-- 100~km/s. A dynamic spectrum, 
shown in Fig.~\ref{Ha_hjd}, was also created from the difference profiles. 
Brighter colours in the plot correspond to enhanced emission and the darker 
colours to the enhanced absorption. The heliocentric Julian dates of the 
observations are shown with crosses in the plot. The data for the times where 
there are no observations are interpolations between the closest timepoints 
with data. The plotting over the heliocentric Julian date instead of the 
rotational phase was chosen, as some events are short lived, and the 
observations in any case cover only slightly more than one rotation. The 
observational phases are given on the left side of the plot.

The most noticeable feature in the dynamical spectrum is the strong enhanced 
absorption around the phase 1.6 at the velocity +100~km/s. If the velocity 
seen in this absorption system was caused by the stellar rotation, it would be 
outside the stellar disk and would not be seen in absorption. Thus it must be 
a cloud of cool gas in the stellar atmosphere that is falling into the star. 
This could be the final stages of a flare event. No evidence of such an event 
is seen in the earlier observations, but it could have occurred during the 
one-night gap in the observations. More enhanced absorption is seen at the 
first observation (phase 0.94) extending to the very blue, to -300~km/s and 
beyond. This could be caused by a mass ejection event with a strong line of 
sight component. In the following observation, more enhanced absorption is seen 
spanning the velocities -100~-- +100~km/s.

Enhanced emission occurs at three main locations: around the phases 1.15--1.30 
at the velocities -40~-- -70~km/s, around the phases 1.2--1.4 at the velocities
+60~-- 100~km/s and at the phases 1.7--1.8 at the velocities +40~-- 70~km/s. 
These features have velocities that place them slightly outside the stellar 
disk, and thus they could be caused by prominences seen at the stellar 
limb. The prominence seen at the blue edge around phases 1.1--1.3 is most 
likely the same one as seen at the red edge 0.5 in a later phase (i.e., at 
phases 1.7--1.8). Also, a weak enhanced absorption feature is seen at phase 
1.4 around the velocity -20~km/s, which could be caused by the prominence 
starting to cross the stellar disk. This prominence could be centred around 
phase 1.5, which in the binary reference frame is the phase pointing away from 
the secondary. The enhanced emission seen in the red around phases 1.2--1.4 
are, based on their velocities, also most likely caused by prominences. 
However, they have to be short lived in nature, as no evidence of them is seen 
in the observations before or after. These prominences would be at the disk 
centre approximately at phases 1.0 and 1.1, which places them on the side 
phasing the secondary. They also coincide with the weaker cool region seen 
around the phases 0.0--0.4 in the Doppler image.

\subsection{Long-term magnetic activity of $\zeta$~And}

The long-term activity in $\zeta$~And is investigated based on the photometric 
$V$ and $y$ band observations obtained with the Wolfgang and Amadeus automatic 
photometric telescopes. The observations between December 1996 and October 2002
were already used in Paper~1. Here, we also use observations obtained between 
June 27, 2003 and October 25, 2008, in total 211 new $V$ magnitudes. 

\begin{figure}
  \centering
  \includegraphics[width=8cm]{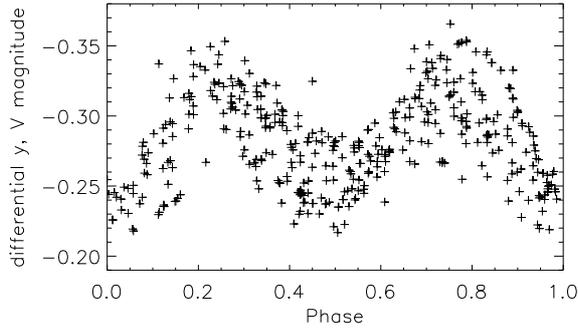}
  \caption{All the differential $V$ and $y$ magnitudes of $\zeta$~And in the 
    instrumental system plotted against the rotational phase. }
  \label{phot_ph}
\end{figure}

\begin{figure}
  \centering
  \includegraphics[width=8cm]{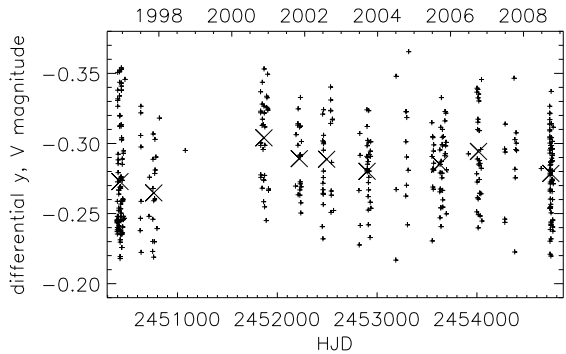}
  \caption{All the differential $V$ and $y$ magnitudes plotted against the 
    heliocentric Julian date. The large crosses show the mean magnitudes for 
    that time. }
  \label{phot_hjd}
\end{figure}

When all the instrumental differential magnitudes are plotted against the 
phase, see plot Fig.~\ref{phot_ph}, the variation caused by the ellipticity 
effect is clearly seen. Still, the observations show much larger scatter 
around the ellipticity curve than is expected from the measurement error of 
0.01--0.02 magnitudes. This indicates that there are also significant 
variations due to starspots. Evidence of changes in the activity level are 
also seen when all the observations are plotted against the Julian Date in 
Fig.~\ref{phot_hjd}. In this plot the small crosses give the individual 
observations and the large crosses the mean of that time period. No mean is 
given for some time periods, as there are so few measurements, or they are 
grouped such, that the full light-curve was not sampled.

Changes in the mean magnitudes could be interpreted as a solar-like activity 
cycle. A spectral analysis of the mean measurements was carried out using the 
Lomb method (Press et al.~\cite{press}). The result indicates a presence of a 
cycle with a cycle length of $5.9\pm 0.7$ years, but the false alarm 
probability is 0.36. Thus the period can be spurious, and more measurements are 
needed for confirming it.

As can be seen from Fig.~\ref{phot_hjd}, $\zeta$~And is on average brighter by 
0.014 magnitudes during the VLT observations presented here than during the 
December 1997 -- January 1998 KPNO observations presented in Paper~1. This 
implies that the spot coverage, and/or the spot temperature, is different 
between the two epochs. For studying this further, the temperature maps 
obtained with the Ca\,{\sc i} 6439~{\AA} line, were investigated both from the 
VLT and KPNO observations. The hottest temperature is basically the same in 
both maps, 4600~K for VLT and 4620~K for KPNO. Still, the coolest temperatures 
are very different. For the VLT observations, the coolest spots are 3710~K and 
for the KNPO map 3480~K. Also, the number of surface elements having a 
temperature of 4000~K or less in the map obtained from the VLT is 75\% of the 
surface elements with those temperatures in the KPNO map. This investigation 
supports the existence of a activity cycle, which is also indicated by the 
long-term photometry of $\zeta$~And. Still, one must keep in mind that the 
temperatures in the Doppler images are very sensitive to the data quality, and 
the VLT data are superior to the KPNO ones.

\section{Conclusions}

The following conclusions can be drawn from the optical interferometry, 
high-resolution spectroscopy and broad band photometry presented in this work.

\begin{enumerate}
\item Optical interferometry gives an apparent diameter of $2.55\pm 0.09$\,mas 
  for $\zeta$\,And. Using the Hipparcos parallax, this translates into a 
  stellar radius of $R=15.9\pm 0.8 {\rm R}_{\odot}$, which is in line with the 
  earlier radius determinations.
\item Combining the interferometrically determined diameter and bolometric 
  flux gives an effective temperature of $T_{\rm eff} = 4665\pm 140$, which is 
  consistent with the values determined through Doppler imaging. 
\item The expected ellipsoidal stellar geometry with $\sim$4\% difference 
  between the long and short axes cannot be confirmed with the current 
  interferometric observations, which have errors of about 4\% in the diameter 
  measurement. However, the highest ellipticity expected for the night 
  of September 18 is consistent with the data.
\item The Doppler images reveal cool spots on the surface of the primary 
  of the $\zeta$~And binary. The spots are located in the equatorial 
  region, and the main concentration of spots is seen around phase 0.75, 
  i.e., 0.25 in phase from the secondary. Another weaker cool region spans the 
  phases 0.0--0.4, again around the equator. There are also indications of a 
  cool polar cap. On the whole, this spot configuration is very similar to the 
  one seen in the earlier published 1997/1998 data.
\item Long-term photometric observations indicate an activity cycle, but more 
  measurements are needed to confirm this and its period. The investigation of 
  the Doppler maps obtained January 1998 and September 2008 also hint at an 
  activity cycle.
\item The chromospheric activity, investigated from the H$\alpha$-line,
  shows evidence of both prominences and cool clouds. The prominences do
  not seem to show any strong evidence of occurring at certain locations in 
  the binary reference frame, nor are they associated with the coolest 
  spot seen on the surface. On the other hand, one of the detected 
  prominences seems to be related to the group of weaker cool spots located at 
  phases 0.0--0.4.
\end{enumerate}

\begin{acknowledgements}
  ZsK is a grantee of the Bolyai J\'anos Fellowship of the Hungarian Academy 
  of Sciences. We also thank the ESO Scientific Visitor Programme for enabling 
  ZsK to visit Garching during the preparation of this paper. This work has 
  made use of the Smithsonian/NASA Astrophysics Data System (ADS) and of the 
  Centre de Donnees astronomiques de Strasbourg (CDS), and the services 
    from the NASA Exoplanet Science Institute, California Institute of 
    Technology, http://nexsci.caltech.edu.
\end{acknowledgements}

\end{document}